\documentclass[ aps, prb, reprint, groupedaddress]{revtex4-1}

\usepackage{graphicx}
\usepackage{amssymb}
\usepackage[squaren]{SIunits}

\usepackage{amsmath}
\usepackage[utf8]{inputenc}	
\usepackage{caption} 
\usepackage[colorlinks=true, linkcolor=blue, citecolor=blue, filecolor=blue, urlcolor=blue, pdfproducer={},pdfcreator={} ]{hyperref}

\usepackage{placeins}
\usepackage{multirow}
\usepackage{braket}

\usepackage{tikz}
\usetikzlibrary{positioning}
\usetikzlibrary{backgrounds}
\usetikzlibrary{snakes}

\newcommand{\ffunc}{\mathcal{F}} 

\makeatletter
\newcommand{\citen}[1]{\def\NAT@spacechar{}[\citenum{#1}]}
\makeatother

\begin{document}

\title{Relaxation of a nonequilibrium phonon distribution induced by femtosecond laser irradiation}

\author{Isabel Klett}

\email{klett@physik.uni-kl.de}

\author{B\"arbel Rethfeld}

\affiliation{Department of Physics and OPTIMAS Research Center, Technical University of Kaiserslautern, Erwin Schroedinger Str. 46, 67663 Kaiserslautern, Germany }


\date{\today}

\begin{abstract}
Ultrafast laser irradiation of solids leads to a thermodynamic nonequilibrium within and between the electron and phonon subsystems of the material.
Due to electron-electron and phonon-phonon collisions, both subsystems relax into respective 
new thermodynamic equilibria within a characteristic thermalization time, which is different for each one of them. Moreover, they equilibrate their temperatures by electron phonon coupling.
The relaxation of the electronic nonequilibrium and its effect on the electron phonon coupling was subject to a number of studies and it is 
comparably well understood, while the nonequilibrium within the phononic subsystem is usually neglected
and its influence of the nonequilibrium phonons on other relaxation processes is unclear. Our calculations show significant differences in the energy transfer rate between the 
electrons and the phonons depending whether a nonequilibrium distribution is assumed for the phonons or not.
Here, we present a model to study the relaxation of the nonequilibrium phonon subsystem. Collisions between phonons are described within the frame of Boltzmann integrals. 
From this, an energy-dependent relaxation time
can be extracted and inserted into a relaxation-time approach. Within the frame of this model, we study the thermalization of a phonon distribution induced by ultrafast laser irradiation.
We show, that the thermalization time of such a distribution is of the order of some hundreds of picoseconds. Moreover, we discuss the energy transfer between Fermi-distributed electrons and nonequilibrium phonons and 
compare this to the energy transfer for equilibrium distributions in both subsystems.
\end{abstract}

\pacs{}

\maketitle

\section{Introduction}
In the past decades, the possibility to generate very short laser pulses of high intensity has emerged. With simple mode-locked laser systems, pulse durations below one picosecond can be achieved. 
By applying more advanced
laser systems, even the generation of laser pulses in the attosecond regime has been reported \citen{Brabec00, Silberberg01}.
There is a wide range of applications for ultrashort laser pulses in research, industry and medicine \citen{Bauerle, Vogel03}.
The option to generate ultrashort laser pulses opens up the opportunity to study microscopic processes in solids on an ultrashort time scale. 
Such processes have been a subject to numerous studies in the past 20 years \citen{Sokolowski98, Cavalleri01, Zijlstra13PRX, Rethfeld02met, Wang94, Sun94, Kaiser00, DelFatti00, Pietanza07, Aoki14, Mueller13PRB}, 
specifically for phonons reaching from 
experimental measurements of coherent phonons \citen{Hase98, Sokolowski03, Harmand13},
to DFT calculations that predict ultrafast phonon bandstructure changes after laser excitation \citen{Klett15, Recoules06, Zijlstra13APA}.



The irradiation of a solid with a femtosecond laser pulse leads to a nonequilibrium state within the electron system as well as between the electron and phonon subsystems, 
due to the fact that the laser energy is mainly absorbed by the electrons, 
while the lattice remains nearly unaffected.
Following the excitation, two main processes occur: (1) The electrons thermalize due to electron-electron collisions, a process that takes place on a timescale of a few up
to several hundreds of femtoseconds \citen{Mueller12,Groeneveld92,DelFatti00, Mueller13PRB}. (2) The electrons
transfer the absorbed laser energy to the lattice due to the electron-phonon coupling. The lattice temperature is rising, while the electrons are cooled down, until a new thermodynamic equilibrium between the 
electron and phonon systems is reached. This process usually takes place on a timescale of a few picoseconds \citen{Groeneveld92}.

A commonly applied minimal model that describes the energy relaxation of electrons and phonons after laser irradiation is the well-known two-temperature model (TTM) \citen{Anisimov74}. 
Within this model, the electrons and phonons are described with different temperatures after laser irradiation, because only the electrons are initially heated by the laser.
The TTM is widely applied to describe the energy relaxation between electrons and phonons \citen{Allen87, Lin08, Hohlfeld00}.
Since this model is based on temperatures, it implicitly assumes equilibrium states within each of both subsystems. 
Therefore,
the application of the TTM on a femtosecond timescale after laser irradiation is questionable.
The limits of the two-temperature model in the description of the relaxation dynamics of electrons and phonons after laser irradiation have been addressed in several 
studies \citen{Groeneveld92, Mueller13PRB, Rethfeld02met, Waldecker16, Carpene06, Kabanov08, Baranov14}.
However, in most of the studies considering the electron relaxation \citen{Rethfeld02met, Pietanza07, Shcheblanov13, Mueller13PRB, Medvedev11} the 
phonon system is assumed to be thermalized. Only recently, the thermalization of the phonon system has been studied by solving a simplified Boltzmann integral \citen{Ono17}.

In \citen{ZimanElPhon} it was shown, that the electrons do not couple to all phonon modes equally. 
For example, in metals, preferably the longitudinal mode absorbs the electron energy, so the assumption of thermalized phonons after 
femtosecond laser irradiation does not hold. The phonons don't absorb the energy of the electrons equally, so, after the exciation of a solid with a femtosecond laser pulse, the phonon subsystem will be in a 
nonequilibrium state. The phonons thermalize due to phonon-phonon collisions, a process, which takes place on a timescale of several picoseconds \citen{Pietanza07}.
The impact of a phonon nonequilibrium on the electron-phonon coupling and the electron distribution is unclear.

In the present work, we study the thermalization of a model nonequilibrium phonon distribution by solving Boltzmann collision integrals, from which we can extract phonon relaxation times. 
These are inserted into a relaxation 
time approach, which gives us insights on the temporal evolution of the phonon distribution. Furthermore, we study the influence of the nonequilibrium phonons on the energy transfer between the electron and 
phonon systems of the 
material.

The paper is organized as follows: First, a model nonequilibrium distribution for phonons after laser irradiation is introduced. Then, we derive the Boltzmann collision integral for the description of phonon-phonon 
processes and apply it to the model nonequilibrium distribution. From the calculated collision term, we extract wavenumber-dependent relaxation times and insert them into a relaxation time approach. This gives an idea 
of the temporal evolution of this nonequilibrium phonon distribution. In the last part of the paper, we check the influence of nonequilibrium phonons on the electronic system by calculating energy transfer rates 
between Fermi-distributed electrons and phonons in nonequilibrium and compare them to energy transfer rates with both subsystems in equilibrium.

\section{Kinetic description}

In this work we describe the change of the distribution functions of electrons and phonons within the frame of the Boltzmann equation.
Although it originates in the kinetic theory of gases, the Boltzmann equation was already successfully applied in the description of quasiparticles in solids \citen{Rethfeld02met, Kaiser00, DelFatti00, Pietanza07}.
We do not consider transport effects and external force fields.
Considering collisions between electrons and phonons and collisions within the respective subsystems, as well as the laser excitation, we end up with the following form of the Boltzmann equation for electrons and phonons
\begin{equation}
\frac{\partial f}{\partial t} =\Gamma_{el-el}+\Gamma_{el-ph}+\Gamma_{abs}^*
\end{equation}
\begin{equation}
\frac{\partial g}{\partial t} =\Gamma_{ph-ph}+\Gamma_{ph-el}+\Gamma_{abs}^* \enspace .
\end{equation}
The terms $\Gamma_i$ hereby represent complete collision integrals. $\Gamma_{el-ph}$ and $\Gamma_{ph-el}$ describe the collisions between electrons and phonons due to the electron-phonon coupling, 
$\Gamma_{el-el}$ denotes the
change of the electron distribution function due to electron-electron collisions, which are responsible for the thermalization of the electronic subsystem. In analogy, the phonon-phonon collisions are represented
by $\Gamma_{ph-ph}$. This collision term describes the thermalization of the phonon system of the material. Finally, $\Gamma_{abs}^*$ stands for the absorption of the laser energy, which enters the electronic system, in 
case that an optical laser pulse was applied. If the frequency of the laser pulse was in the terahertz regime, it would be possible to exciten the phonons directly \citen{Manceau10}.
Most of these collison integrals are well known and have already been applied in the description of the temporal evolution of the electron and phonon system after femtosecond laser irradiation 
\citen{Rethfeld02met, Kaiser00, Mueller13PRB, Pietanza07}. 
The phonon-phonon collision term so far was only calculated with severe simplifications \citen{Ono17} in the phonon-phonon matrix element.
This study will concentrate on this phonon-phonon collision term, which is derived below. 

We apply our collision term to an assumed nonequilibrium distribution of the phonons after laser excitation. Such distributions were calculated for instance in Ref \citen{Kaiser00}.


 \begin{figure}[!h]
\includegraphics[width=0.5\textwidth]{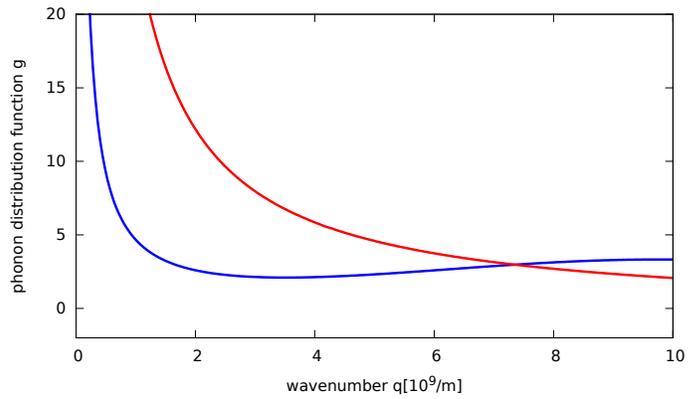} 
\caption{Model of an examplary nonequilibrium phonon distribution after excitation by hot electrons (blue) and a Bose distribution of equal internal energy (red). }
\label{abb:distribution}
\end{figure}

Figure \ref{abb:distribution} shows the applied nonequilibrium distribution and a Bose distribution 
\begin{equation}
g(q,T)=\frac{1}{e^{\frac{\hbar q c_s}{k_BT}}-1} \enspace .
\end{equation}
of the same internal energy. For small wavenumbers $q$, the distribution resembles a Bose distribution of 
lower internal energy than the nonequilibrium distribution, for larger wavenumbers, the occupation 
number is higher. A similar trend of the nonequilibrium distribution for phonons after laser irradiation was calculated in Ref \citen{Kaiser00}.
The internal energy of the Bose distribution corresponds to a temperature of 1634 Kelvin, that is slightly lower than the melting temperature of silicon.



\section{Phonon-phonon collisions}
This study deals with the relaxation of nonequilibrium phonons. In the following, we present the derivation of the phonon-phonon collision term $\Gamma_{ph-ph}$.

For many studies, isotropic collision terms can be assumed \citen{Mueller13PRB, Pietanza07}.
However, this assumption is insufficient for the description of phonon-phonon collisions. The phonon-phonon collision term has to be kept direction-dependent, in order to distinguish between the different 
phonon modes.
Phonon-specific quantities like frequencies and polarization vectors follow from the chosen interatomic potential and the crystal structure of the material. These quantities are material specific and enter the matrix
element for phonon-phonon collisions.

\begin{figure}
\centering
\begin{tikzpicture}
\draw[->, ultra thick] (0,0) node[anchor= north east] {\large $-\vec q$} -- (1.5,0) ; 
\draw[->, ultra thick] (1.5,0) -- (2.5,1) node[anchor= south west] {\large $\vec {q}\, '$}; 
\draw[->, ultra thick] (1.5,0) -- (2.5,-1) node[anchor= north west] {\large $\vec {q}\, ''$}; 

\draw[->, ultra thick] (5,1) node[anchor= south east] {\large $-\vec q$} -- (6,0) ; 
\draw[->, ultra thick] (5,-1) node[anchor= north east] {\large $-\vec {q} \, '$} -- (6,0) ; 
\draw[->, ultra thick] (6,0) -- (7.5,0) node[anchor= west] {\large $\vec {q} \, ''$} ;  
\end{tikzpicture}
\caption{Possible interaction processes between three phonons. Either one phonon decays into two others or two phonons combine into one.}
\label{abb:phphcoll}
\end{figure}
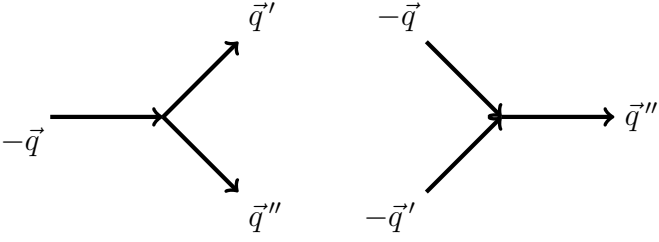
The most simple interaction processes between phonons are the creation or annihilation of one phonon. As shown in figure \ref{abb:phphcoll},
both of these processes contain three particles. In the creation process one phonon decays into two other phonons,
in the annihilation process two phonons combine into one. $\vec q$ hereby denote the wavevectors of the participating phonons, $\vec q \,'$ and $\vec q \,''$ stand for the collision partners.
To describe these processes, the harmonic approximation for the Hamiltonian is not sufficient, instead a Taylor expansion of the effective
potential up to the third order is required. Therefore, the chosen interatomic potential has to be at least a three-body potential. In our case, we apply a Stillinger-Weber potential \citen{Stillinger85}.
The Hamiltonian for phonon-phonon interactions denotes as \citen{ZimanElPhon}
\begin{align}
H_{ph-ph}&= \frac{1}{3!} \sum_{\vec{\ell}, \vec b,\vec{\ell}',\vec b',\vec{\ell}'',\vec b''} \sum_{\alpha , \beta , \gamma} r_{\vec{\ell} \vec b}^{\alpha} r_{\vec{\ell}'\vec b'}^{\beta} r_{\vec{\ell}''\vec b''}^{\gamma} 
\nonumber \\& \cdot \left(\frac{\partial^3 V}{\partial \vec r_{\vec{\ell} \vec b} \partial \vec r_{\vec{\ell}'\vec b'} \partial \vec r_{\vec{\ell}'' \vec b''}}\right)_{ \alpha \beta \gamma}  \enspace ,
\end{align}
where the $\vec r_{\vec{\ell}\vec b}$ represent the displacement vectors of the atoms out of their equilibrium position. The vectors $\vec{\ell}$ hereby denote vectors to the origins of the unit cells, 
$\vec{b}$ the vectors from 
the origin of the unit cell to the different atoms. The primed quantities
refer to the respective displaced atoms.
The term contains a triple inner product between the displacement vectors and the tensor of the third derivative with respect to the interatomic potential to the displacement vectors written in component notation
as sum over $\alpha, \beta, \gamma$.
Following some transformations \citen{ZimanElPhon}, the phonon-phonon interaction Hamiltonian can be rewritten as a term consisting of creation and annihilation operators
\begin{align}
H_{ph-ph} &= \frac{1}{3!} \sum_{\vec{q},p,\vec{q}',p',\vec{q}'',p''} \delta_{\vec{G},\vec{q}+\vec{q}'+\vec{q}''}  F_{\alpha \beta \gamma} (a_{\vec{q}p}^{+}-a_{-\vec{q}p}) 
\nonumber \\& \cdot (a_{\vec{q}'p'}^{+}-a_{-\vec{q}'p'}) (a_{\vec{q}''p''}^{+}-a_{-\vec{q}''p''}) \ffunc_{\vec{q}p\vec{q}'p'\vec{q}''p''} \enspace ,
\end{align}
with $\vec G$ denoting a reciprocal lattice vector, $p$ denoting the phonon modes, the primed quantities hereby denote the modes of the collision partners, and
\begin{align}
\ffunc_{\vec{q}p\vec{q}'p'\vec{q}''p''} &= i (\frac{1}{2} \hbar)^{3/2} (NV_K)^{-1/2} (\nu_{\vec{q}p} \nu_{\vec{q}'p'} \nu_{\vec{q}''p''})^{-1/2} 
\nonumber \\& \cdot \sum_{\vec{b},\vec{b}',\vec{b}''} (m_{\vec{b}} m_{\vec{b}'} m_{\vec{b}''})^{-1/2} 
 \sum_{\alpha, \beta, \gamma} {\textbf{e}}_{\vec{q}\vec{b}p}^{\alpha} {\textbf{e}}_{\vec{q}'\vec{b}'p'}^{\beta} {\textbf{e}}_{\vec{q}''\vec{b}''p''}^{\gamma} 
\nonumber \\& \cdot \sum_{\vec h',\vec h''} e^{-i\vec q'\vec h'}e^{-i\vec q''\vec h''} \left(\frac{\partial^3 V}
{\partial \vec r_{\vec{\ell} \vec b} \partial \vec r_{\vec{\ell}+\vec h',\vec b'} \partial \vec r_{\vec{\ell}+\vec h'', \vec b''}}\right)_{ \alpha \beta \gamma}  \enspace  .
\end{align}
This Hamiltonian contains the triple inner product of the polarization vectors $\vec {\bf e}$ with the tensor of the third derivatives of the interatomic potential with respect to the atom displacement vectors.
Additionally, we define the vectors $\vec{h}'$ and $\vec{h}''$ as $\vec{\ell}' = \vec{\ell} +\vec{h}'$ and $\vec{\ell}''= \vec{\ell} +\vec{h}''$.
The $\nu$ denote the phonon frequencies, the $m$ are the atomic masses and $NV_K$ is the product of the considered crystal volume and the number of unit cells per volume.
The phonon frequencies and polarisation vectors were calculated from the eigenvalues and eigenvectors of the dynamical matrix.

\subsection{The Phonon-Phonon collision term}

With the Hamiltonian for three phonon interactions derived above, Fermi's golden rule can be applied to derive a Boltzmann collision term for phonon-phonon scattering processes
leading to a collision term
\begin{equation}
\Gamma_{ph-ph}=\frac{2\pi}{\hbar}|\bra 1  H_{ph-ph} \ket 0|^2 \delta (E_1 - E_0) \enspace .
\end{equation}

To evaluate this expression, the commutation relations valid for phonons
\begin{equation}
[a_i,a_j^+]=\delta_{ij} \enspace , \enspace [a_i^+,a_j^+]=0 
\end{equation}
are applied.
Additionally, the following terms for creation and annihilation operators apply in the case of phonons
\begin{equation}
<a_i^+a_i>=g(E_i) \enspace ,
\end{equation}
and
\begin{equation}
<a_i a_i^+>=(1+g(E_i)) \enspace ,
\end{equation}
with the phonon distribution function $g$ at given energy $E_i$, with $E=\hbar c_s q$.
Taking all these relations into account, a Boltzmann collision term for phonon-phonon interactions is calculated

 \begin{align}
	&\Gamma_{\text{ph-ph}}(\vec{q}) = \frac{2 \pi}{\hbar} \left( \frac{1}{3!} \right)^2 \sum_{p',p''} \sum_{\vec{q}',\vec{q}''} \delta_{\vec{G},\vec{q}+\vec{q}'+\vec{q}''}  | M |^2  \mathcal G 
      \enspace \delta_{E_1, E_0}  \enspace ,
      \label{eq:collterm}
\end{align}
with
\begin{align}
	M &=  \enspace i \left( \frac{1}{2} \hbar \right)^{3/2} (NV_K)^{-1/2}    
  \sum_{b,b',b''} (m_{\vec{b}} m_{\vec{b}'} m_{\vec{b}''})^{-1/2}  
 \nonumber     \\ & \cdot {(\nu_{\vec{q}p} \nu_{\vec{q}'p'} \nu_{\vec{q}''p''})^{-1/2}} 
    \sum_{\alpha, \beta, \gamma} {{{\textbf{e}}_{\vec{q}\vec{b}p}^{\alpha} {\textbf{e}}_{\vec{q}'\vec{b}'p'}^{\beta} {\textbf{e}}_{\vec{q}''\vec{b}''p''}^{\gamma}}} 
 \nonumber     \\ & \cdot \sum_{\vec{h},\vec{h}'} e^{-i\vec{q}'\vec{h}'}e^{-i\vec{q}''\vec{h}''} { \left( \frac{\partial^3 V}{\partial r_{\ell b} \partial r_{\vec{\ell}+\vec{h}',\vec{b}'} \partial r_{\vec{\ell}+\vec{h}'', \vec{b}''}} \right)_{ \alpha \beta \gamma}} \enspace ,
\end{align}
and 
\begin{align}
\mathcal G=(g+1)g_{-}'(g''+1)-g_{-}(g'+1)(g''+1)-g_{-}g_{-}'(g''+1) 
\nonumber \\+(g+1)(g'+1)g_{-}''+(g+1)g_{-}'g_{-}''-g_{-}(g'+1)g_{-}'' \enspace .
\end{align}
The functional $\mathcal G$ represents a term over the distribution functions of the phonons, indicating the gain and loss of phonons in the different states. The abbreviation $g$ here refers to the value of the 
distribution $g(E(\vec q))$ at the given state.

The phonon-phonon collision term contains multiple sums, two are sums over the wave vectors $\vec q$ of the scattered phonons, the others run over the modes $p$, respectively.
The sums over the wave vectors can be rewritten in integral form leading to six integrals over the components of the wave vectors.
By means of an evaluation of the delta functions representing energy and momentum conservation, the number of integrals required can be reduced to two.

%
 \begin{figure}[!h]

\includegraphics[width=0.5\textwidth]{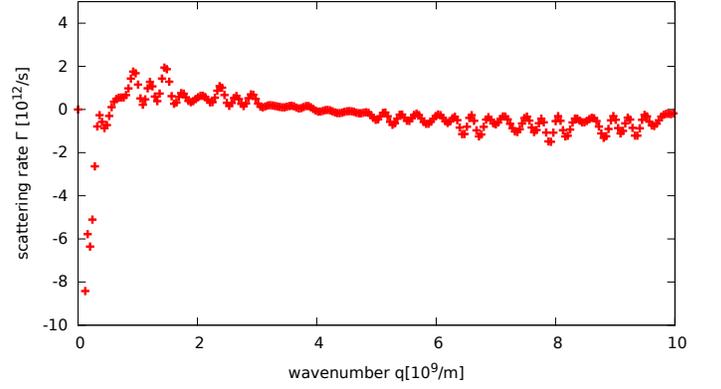} 

\caption{Collision term $\Gamma_{ph-ph}$ according to Equation \eqref{eq:collterm} of the assumed nonequilibrium phonon distribution function, shown in Figure \ref{abb:distribution} .}
\label{abb:gamma}
\end{figure}

In figure \ref{abb:gamma}, the temporal derivative of the assumed nonequilibrium phonon distribution (see Figure \ref{abb:distribution}) is shown.

For the calculation we took a material with an fcc structure, so only acoustic phonons are possible. We applied material parameters of silicon, the excitend phonon mode is the longitudinal one.
We apply a Debye temperature of 645 Kelvin, a sound velocity of 8433 $\frac{m}{s}$ an atomic mass of 28.0855 u and a lattice parameter of 5.431 A. We consider atomic displacements and phonon wavevectors in 
(111) direction.

\subsection{Temporal evolution of the phonon distribution}

Calculating the complete temporal evolution of the phonon system with the Boltzmann collision integral would be extremely time-consuming. A compromise would be a relaxation time approach.
The assumption hereby is, that a system evolves towards an equilibrium state due to collision processes.
Within this approximation, the time derivative of the distribution function writes as following
\begin{equation}
\Gamma_{ph-ph}= \frac{\partial g(q)}{\partial t} = \frac{g(q)-g_{Bose}(q)}{\tau(q)} \enspace .
\label{eq:relaxationapp}
\end{equation}
The change of the nonequilibrium distribution $g$ is thus given by its difference to a Bose distribution of equal internal energy divided by the relaxation time $\tau$. 
All quantities depend on the phonon wave number $q$, and therefore also on the phonon energy.

Since we determine the time derivative of the phonon distribution $g(q)$ through the Boltzmann collision integral, we can extract relaxation times, which depend on the wavenumber $q$
\begin{equation}
  \tau(q)= \left |\frac{g(q)-g_{Bose}(q)}{\Gamma_{ph-ph}} \right | \enspace .
  \label{eq:relaxationtimes}
\end{equation}

A problem of this approach is that the internal energy is not exactly conserved when applying wave-number-dependent relaxation times. Therefore, we checked in the following calculations, that these deviations are 
relatively small. The highest deviation in the internal energy is found to be around 9 \%. 

\begin{figure}[!h]
\includegraphics[width=0.5\textwidth]{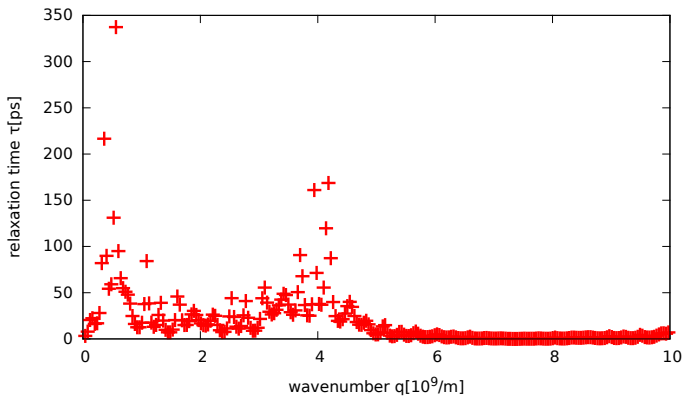} 
\caption{Calculated phonon relaxation times depending on the wavenumber $q$.}
\label{abb:rel}
\end{figure}

The relaxation times of the nonequilibrium phonon distribution shown in figure \ref{abb:distribution} after laser irradiation were calculated through equation \eqref{eq:relaxationtimes} and are shown in figure 
\ref{abb:rel}. 
The relaxation times for small wave numbers are on a timescale of up to
some hundreds of picoseconds, while for higher wavenumbers, the relaxation times are much smaller. These results indicate that the phonon distribution thermalizes on timescales of a few hundred picoseconds.

The wavenumber-dependent relaxation times can be inserted into a relaxation time approach \eqref{eq:relaxationtimes} and integrated in time. This gives an idea of the temporal evolution of the phonon distribution.

\begin{figure}[!h]
\includegraphics[width=0.5\textwidth]{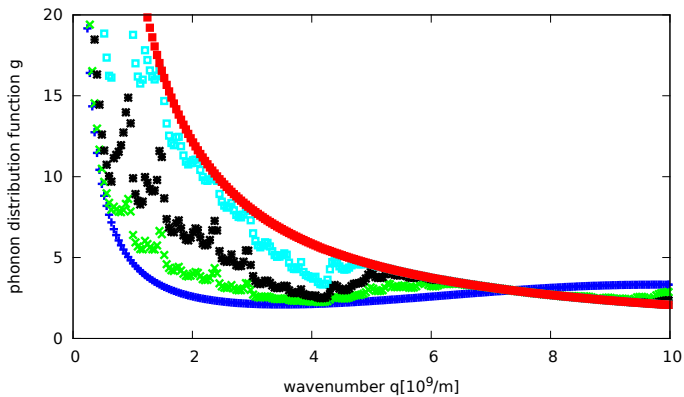} 
\caption{Relaxation of a phonon nonequilibrium distribution. The plotted timesteps are the initial nonequilibrium distribution (blue), the distribution after 3 ps (green), 10 ps (black), 40 ps (light blue) and
the relaxed distribution after 500 ps (red).}
\label{abb:relaxation}
\end{figure}

In figure \ref{abb:relaxation}, such temporal evolution of the nonequilibrium distribution function within 500 ps is plotted. A comparatively fast relaxation of the 
nonequilibrium distribution at higher wavenumbers can be observed, where it is 
already almost thermalized after 3 picoseconds,
while the thermalization of the whole distribution function to a new equilibrium distribution takes much longer. After 40 picoseconds, the distribution still deviates from an equilibrium distribution, particularly at 
small wave numbers.
This indicates that processes driven by phonons of small energy are influenced by the phononic nonequilibrium on timescales in the 100 ps range.

\section{Influence of a phonon nonequilibrium on the electron-phonon energy transfer}

It is interesting to check how strong the influence of a nonequilibrium distribution of the phonons
on the electron system and on the coupling between electrons and phonons is.
It was already shown, that nonequilibrium electrons influence the electron phonon coupling \citen{Rethfeld02met,Mueller13PRB,Groeneveld92,Giri15}. 
Therefore, it can be expected, that nonequilibrium phonons also affect the energy relaxation between electrons and phonons.
In this section we apply the model presented in \citen{Brouwer17} to determine the strength of the electron-phonon coupling. We assume material parameters as given in \citen{Brouwer17}, but apply the nonequilibrium 
distribution discussed before (see Fig \ref{abb:distribution}) as the starting distribution for the phonons.

In this calculation, only electron-phonon interactions are considered, all other interaction processes are switched off.
We calculate the energy transfer rate from the electrons to the phonons according to equation (10) in \citen{Brouwer17} applying the phonon nonequilibrium distribution as starting distribution while the electrons are 
Fermi distributed 
at 10000 Kelvin at the beginning of our calculation. For comparison, the energy transfer rate is also determined assuming equilibrium distributions for both subsystems. In that calculation, at each timestep the 
distributions of electrons as well as of phonons are set to their corresponding equilibrium distributions of the same internal energy.

 \begin{figure}[!h]
 \centering
 \begin{minipage}{0.5\textwidth}
\includegraphics[width=\textwidth]{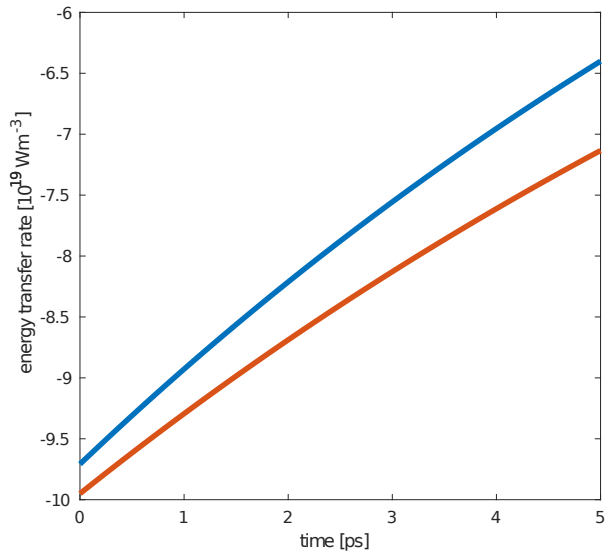} 
\end{minipage}
\caption{Rate of the energy transfer from the electrons to the phonons assuming a phonon nonequilibrium (blue) and a Bose distribution at the same internal energy (red). }
\label{abb:etransfer}
\end{figure}

We observe a considerable difference in the energy transfer rate from the electrons to the phonons depending on the phonon nonequilibrium, as shown in figure \ref{abb:etransfer}.
This difference indicates that the phonon nonequilibrium also plays an important role in the evolution of the electron system and the electron-phonon coupling and should therefore not simply be neglected.
As a consequence, it is further important to identify a timescale, after which the phonons are thermalized and the assumption of a Bose distribution is valid.

\begin{figure}[!h]
\includegraphics[width=0.5\textwidth]{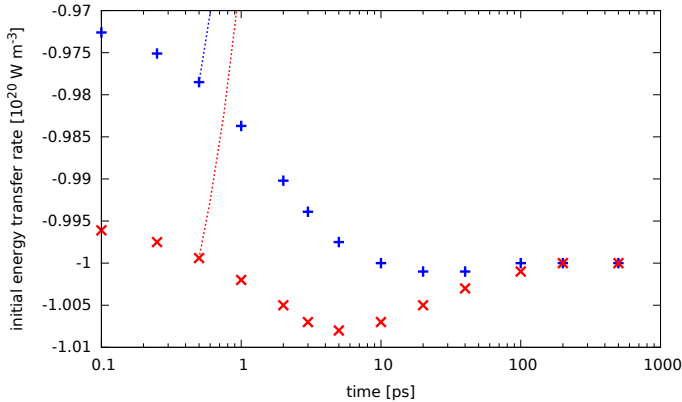} 
\caption{Rate of the energy transfer from Fermi-distributed electrons to the nonequilibrium phonons at different timesteps (blue) and Bose-distributed phonons with the same internal energy (red).}
\label{abb:etrans_z}
\end{figure}

Therefore, the initial energy transfer rates between Fermi distributed electrons and nonequilibrium phonons in different stages of the relaxation process are calculated. We determine the electron-phonon coupling 
between electrons at 10000 Kelvin and phonons at different nonequilibrium states, calculated according to equation \eqref{eq:relaxationtimes} and shown for some delays in Figure \ref{abb:relaxation}. With these 
initial conditions, initial energy transfer rates are calculated. The results are shown in comparison to the coupling to equilibrium phonons of the same respective energy in Figure \ref{abb:etrans_z}.
The blue dots denote the energy transfer rate from the electrons to the phonons in the nonequilibrium distribution at certain timesteps during the relaxation process,
the red dots are the energy 
transfer rates
calculated for 
a Bose distribution of the same respective internal energy. The electrons are always Fermi distributed at 10000 Kelvin. The dashed lines represent the temporal evolution of the energy transfer rate for a 
certain time, similar to the curves in
Figure \ref{abb:etransfer}.
The energy transfer from the electrons to the phonons is decreasing over time for the phonons in different 
stages of the thermalization process and the rates for the nonequilibrium distributed phonons are approaching the rates for the Bose distributed phonons, until there is no more observable difference at 200 ps.
This result indicates, that the assumption of Bose distributed phonons would not make a difference on the electron phonon coupling anymore from this point on.
Note that the minimum in the red curve is an artifact due to deviation of the internal energy of the phonons during the calculated relaxation process.

\section{Conclusion}

The thermalization of the phonons is studied within the frame of a Boltzmann collision term, which gives a wavenumber-dependent relaxation time. We found these relaxation times to be
on a timescale of up to some hundreds of picoseconds. The temporal evolution of a model nonequilibrium phonon distribution after laser irradiation is studied in the framework of a relaxation time approximation. 
A comparison of the initial energy transfer rates between electrons and phonons in nonequilibrium and thermalized indicates
that the thermalization process of the given phonon distribution takes place on a timescale of a few hundred picoseconds.
Additionally, we study the energy transfer between Fermi-distributed electrons and nonequilibrium phonons, which is compared to the energy transfer for equilibrium distributions in both subsystems. A comparison of 
the initial energy transfer rates in both cases indicates that the thermalization process of the given phonon distribution takes place on a timescale of a few hundred picoseconds.
In conclusion, we have observed a considerable influence of a thermodynamic nonequilibrium of the phonons on the energy transfer between the electron and phonon systems.

\section{Acknowledgements}
The authors thank Nils Brouwer for providing the electron-phonon code. Financial support of the former MBWWK Rheinland-Pfalz, the TU-Nachwuchsring and Deutsche Forschungsgemeinschaft (grant RE 1141/11-1 and RE 1141/15-1)
is gratefully acknowledged.

\FloatBarrier
\bibliographystyle{unsrt}

\bibliography{/home/isabel/latexkrams/Paper/paper_neqph.bib}

\end{document}